\documentclass[12pt,a4paper]{article}
\begin{document}
\title{QUANTUM MODEL OF PHONON TRANSPORT AND HEAT CONDUCTIVITY IN CARBON 
NANOCLUSTERS AND NANOTUBES }
\maketitle

 E.Ya.Glushko (1 and 2), V.N.Evteev (3), M.V.Moiseenko (3),
 N.A.Slusarenko (3) and A.A.Zakhidov (4)

((1) Institute for Semiconductors Physics NAS Ukraine, Pr.Nauki. 21,
Kiev, 03047, Ukraine;
(2) Slavonic University, Anry Barbusse Str. 9, Kiev  03150, Ukraine;
(3) Pedagogical University, Gagarin Avenue 54, Krivoy Rog, 50086, Ukraine;
(4) Texas University at Dallas, Richardson, TX 75083-0688, USA)

(E-mail: eugene.glushko@scientist.com)
\begin{abstract}
   A complex approach phonon quantum discrete model (PQDM) was developed to 
describe dynamics, kinetics and statistics of phonons in carbon nanostructures 
with zero-chirality of both zig-zag and armchair geometry. The model allows 
include into the pure phonon problem existing interaction with others 
subsystems: electrons, photons, impurities and defects. We predict that planar 
C- structures are geometrically stable and may bridge interelectrode space in 
strong external electric field. The exact solution of generalized thermal 
conductivity (TC) equation was obtained for nanotubes. Temperature distribution 
along the tube was derived analytically. The diagonalization procedure for the 
case ofstrong ph-ph interaction was proposed. It was shown the quadratic 
increasing of heat conductivity with the growth of the phonon mean free path 
(PMFP). Heat capacitance and the entropy of carbon linear tubes were calculated 
as the function of temperature. Our theoretical approach explains the nature of 
good TC in carbon and carbon-like materials by existing of the soft vibration 
branch (low frequency radial breathing mode phonons with high density of states 
at thermal energies) accompanied by structure hardness (high frequency 
$\varphi$- and z-branches) providing big PMFP. TC coefficient for high 
conducting channel in surrounding medium was calculated. The mechanism of heat 
conductivity temperature damping was analyzed. Two competitive tendencies 
produce TC maximum at intermediate temperatures (100-300)K. It was shown the 
strongly non-linear increasing of effective heat conductivity with the growth 
nanotubes concentration. It was shown that insertion of armchair nanotube 
inside a medium or its coating by polyacetilene molecule considerably changes 
the structure of radial breathing phonons.
\end{abstract}

\newpage
I. Motivation

II. Introduction

III. Systems under consideration

IV. Phonon quantum discrete model (PQDM)

V. Vibrational dynamics. Linear approximation.

VI. Frequency spectrum and DOS. Zig-zag geometry.

VII.  Frequency spectrum and DOS. Armchair geometry. 

VIII. Amplitude distribution and knot's theorem
IX. Heat transfer
X. Phonon mean free path and size dependences
XI. Thermal coefficient temperature dependence
XII.  Generalized phonon transport equation

XIII. Summary and discussion

\section{ Motivation}

 Phonon dynamics in open and closed carbon and carbon-like nets. How many phonon 
branches are present? Whether exists a simple correlation between differed 
phonon types for different NT geometries?

 Vibrational eigenmode and its amplitudes distribution in the net.

 Vibrational eigenmodes, phonons, micro-sound and heat propagation. Phonon mean 
free path. Sound waves boundary conditions.

 The mechanism of temperature damping at high temperatures and concrete law of 
increasing at low temperatures. 

 Maximal thermal conductivity for solitary carbon tube is not measured reliably 
now and changes in wide interval from 200 W/mK to 3000 W/mK for different 
authors.

 The temperature of maximal thermal conductivity for carbon tubes is not surely 
measured also and may change in wide interval from 150 K to 300 K for different 
authors. 

 The question is whether exist differences in thermal conductivity of solitary 
carbon tubes depending on type (armchair or zig-zag), chirality and diameter.
 Which physical factors are essential in electron transport through the 
molecule?

 Do phonons (vibrons) participate essentially in electron transport through the 
molecule? 

 Are differences in thermal conductivity of solitary carbon tubes, flat carbon 
structures like graphene and graphite (in plane) essential? 
\newpage
\section{ INTRODUCTION}
\special{em:graph fig_2_1.gif}
\vspace{24.0cm}
{ Fig.2.1 General picture of connected subjects for the problem of phonon 
propagation in molecular nets. }
\newpage
\section{  Systems under consideration }

\special{em:graph fig_3_1.gif}
\vspace{4.47cm}
{ Fig.3.1  A polymer molecule absorbed by armchair nanotube. Absorption bonds 
are shown by red lines, C-C bonds of the polymer are shown by violet color. }

\vspace{0.5cm}
\special{em:graph fig_3_2.gif}
\vspace{4.39cm}
{Fig.3.2 Heat transistor (3-polar contact) on a flat carbon structure connecting 
three baths at different temperatures $T_l$ , $T_u$ and $T_r$.  }

\section{  Phonon quantum discrete model (PQDM) } 

\begin{eqnarray}
\mathop {H_0}\limits^{\wedge}   = \sum\limits_{s,\sigma } {\hbar \omega 
_{s\sigma } 
 (\hat b_{s\sigma }^ +  \hat b_{s\sigma }  + 1/2)} 
\label{eq4.1}
\end{eqnarray}

\begin{eqnarray}
\begin{array}{l}
\mathop H\limits^ \wedge  _{{\mathop{\rm int}} }  = 
\sum\limits_{i_l ,\sigma } {G_{i_l \sigma } (\hat b_{i_l \sigma }^ +  
\hat b_l  + \hat b_l^ +  \hat b_{i_l \sigma }^{} )}  +  \\ 
+ \sum\limits_{i_r ,\sigma } {G_{i_r \sigma } (\hat b_{i_r \sigma }^ +  
\hat b_r  + \hat b_r^ +  \hat b_{i_r \sigma }^{} )}  \\ 
 \end{array}
\label{eq4.2}
\end{eqnarray}

\section{ Vibrational dynamics. Linear approximation.}
  Zig-zag NT 

\begin{eqnarray}
{m\ddot \rho _i  =  - k'(3\rho _i  - \rho _{i1}  - \rho _{i2}  - 
\rho _{i3} )} 
\label{eq5.1}
\end{eqnarray}

\begin{eqnarray}
 {m\ddot z_i  =  - k(z_i  - z_{i1} ) - 0.25k(2z_i  - z_{i2}  - z_{i3} )} 
\label{eq5.2}
\end{eqnarray}

\begin{eqnarray}
 {m\ddot x_i  =  - 0.75k(2x_i  - x_{i1}  - x_{i2} ) - k'(x_i  - x_{i3} )} 
\label{eq5.3}
\end{eqnarray}

  Armchair NT 

\begin{eqnarray}
 {m\ddot z_i  =  - 0.75k(2z_i  - z_{i1}  - z_{i2} ) - k'(z_i  - z_{i3} )} 
\label{eq5.4}
\end{eqnarray}

\begin{eqnarray}
{m\ddot x_i  =  - k(x_i  - x_{i1} ) - 0.25k(2x_i  - x_{i2}  - x_{i3} )} 
\label{eq5.5}
\end{eqnarray}

\begin{eqnarray}
\hat D\vec A = \omega ^2 \vec A
\label{eq5.6}
\end{eqnarray}

\begin{eqnarray}
\omega _{\sigma s} ,\vec A_{\sigma s}^ +   = 
(C_{s1} ,C_{s2} ,C_{s3} ...C_{sr} ),r = 2n(m + 1)
\label{eq5.7}
\end{eqnarray}

\special{em:graph fig_5_1.gif}
\vspace{4.91cm}
{ Fig.5.1 The problem's geometry. Z-axes is perpendicular to the figure plane. 
It's shown j and r shifts of an atom. }

\special{em:graph fig_5_2.gif}
\vspace{3.12cm}
{ Fig.5.2. The problem's geometry. F is z-axes projection of the force acting 
between a and b atoms. dz marks the atom a shift along the axes z, dzcos(p/3) is 
the shift projection.  }

\newpage
.

\special{em:graph fig_5_3.gif}
\vspace{9.04cm}
{ Fig.5.3. Sketch of the dynamical matrix for zig-zag tubulene created from a 
graphene sheet \{5,5\} containing 60 atoms by its rolling up. Rose-colored spots 
are diagonal elements, green and blue show two types of bonds, rest elements 
equal to zero, left and upper red bands contain atomic numbers.  }

\vspace{0.5cm}
\special{em:graph fig_5_4.gif}
\vspace{9.15cm}
{ Fig.5.4. Sketch of the dynamical matrix for armchair tubulene created from a 
graphene sheet \{5,5\} containing 60 atoms by its rolling up. Rose-colored spots 
are diagonal elements, green and blue show two types of bonds, rest elements 
equal to zero, left and upper red bands contain atomic numbers.  }

\section{Frequency spectrum and DOS. Zig-zag geometry.} 

\begin{eqnarray}
g(\omega _{s\sigma } ) \approx {1 \over {\omega _{s\sigma }  - 
\omega _{s - 1,\sigma } }}
\label{eq}
\end{eqnarray}

\vspace{0.5cm}
\special{em:graph fig_6_1.gif}
\vspace{3.49cm}
{ Fig.6.1. Radial mode spectrum of zig-zag tubulene created from a graphene 
sheet \{5,15\} by rolling up around z-axes. The unity of frequency is $\omega_0$   
}

\vspace{0.5cm}
\special{em:graph fig_6_2.gif}
\vspace{4.99cm}
{ Fig.6.2 Radial mode DOS function for Fig.6.1 system. h=Int[35r/50].   }

\vspace{0.5cm}
\special{em:graph fig_6_3.gif}
\vspace{4.28cm}
{ Fig.6.3 Azimutb mode spectrum of zig-zag tubulene created from a graphene 
sheet \{5,15\} by rolling up around z-axes. The unity of frequency is $\omega_0$ 
.  }

\newpage
.

\special{em:graph fig_6_4.gif}
\vspace{4.62cm}
{ Fig.6.4 Azimuthal mode DOS function (arbitrary units) for graphene sheet 
\{5,15\} h=Int[35r/50].   }

\vspace{0.5cm}
\special{em:graph fig_6_5.gif}
\vspace{5.10cm}
{ Fig.6.5 Axial mode spectrum of zig-zag tubulene created from a graphene sheet 
\{5,15\} by rolling up around z-axes. The unity of frequency is $\omega_0$ .  }

\vspace{0.5cm}
\special{em:graph fig_6_6.gif}
\vspace{6.16cm}
{ Fig.6.6 Axial mode DOS function (arbitrary units) for tubulene fragment 
created from a graphene sheet \{5,15\} by rolling around z-axes. Horizontal 
axes, frequency in $\omega_0$  }

\newpage
\section{Frequency spectrum and DOS. Armchair geometry. } 

\vspace{0.5cm}
\special{em:graph fig_7_1.gif}
\vspace{11.19cm}
{ Fig.7.1. Radial mode density of states for armchair tubulene created from a 
graphene sheet \{8,15\} by rolling up around z-axes. Insertion the same for zig-
zag tubulene.  }

\newpage
.

\special{em:graph fig_7_2.gif}
\vspace{10.82cm}
{ Fig.7.2 Tangential j-mode density of states for armchair tubulene created from 
a graphene sheet \{8,15\} by rolling up around z-axes. Insertion is zig-zag 
equivalent for this that is z-mode DOS. h=Int[4r/5].  }

\vspace{0.5cm}
\special{em:graph fig_7_3.gif}
\vspace{8.01cm}
{ Fig.7.3 Axial z-branch density of states for armchair tubulene created from a 
graphene sheet \{8,15\} by rolling up around z-axes. Insertion is zig-zag 
equivalent for this that is $\varphi$-mode DOS.  }

\newpage
.

\special{em:graph fig_7_4.gif}
\vspace{7.35cm}
{ Fig.7.4 Lower part. Three branches of phonon spectrum for armchair tubulene 
created from a graphene sheet \{8,15\} by rolling up around z-axes. Upper part. 
Radial phonon band for zig-zag \{8,15\} NT.  }

\section{Amplitude distribution and knot's theorem } 

\begin{eqnarray}
\mathop {A_j }\limits^{}  = \sum\limits_{s,\sigma } 
{\left| {C_{sj} } \right|^2 n_{s\sigma } } ,
\label{eq}
\end{eqnarray}

\vspace{0.5cm}
\special{em:graph fig_8_1.gif}
\vspace{6.35cm}
{ Fig.8.1 Calculated distribution of vibrational amplitudes of \{8,8\} planar 
carbon net for the 4-th mode.  }

\newpage
.

\special{em:graph fig_8_2.gif}
\vspace{3.35cm}
{ Fig.8.2. Calculated r-branch amplitude distribution s=5 along armchair 
tubulene created from a graphene sheet \{6,5\} by rolling up around marked 
direction. Transversal knot's lines are degenerated. }

\vspace{0.5cm}
\special{em:graph fig_8_3.gif}
\vspace{4.04cm}
{ Fig.8.3 r-branch amplitude distribution along zig-zag tubulene created from a 
graphene sheet \{6,5\} by rolling up around marked direction. 5th state. 
Longitudinal knot's lines are degenerated. }

\vspace{0.5cm}
\special{em:graph fig_8_4.gif}
\vspace{2.48cm}
{ Fig.8.4 Calculated mean square amplitude distribution along zig-zag tubulene 
created from a graphene sheet {15,5} by rolling up around marked direction. 
T=0.03 eV. Circular arrows show rolling up of the structure.  }

\section{ Heat transfer }   

\begin{eqnarray}
\dot W_{l \to s\sigma }  = {{2\pi } \over \hbar }\left| {G_{ls} } 
\right|^2 g_l (\omega _{s\sigma } )N_l (\omega _{s\sigma } )
(1 + n(\omega _{s\sigma } ))
\label{eq9.1}
\end{eqnarray}

\begin{eqnarray}
n_s  = {{|\sum\limits_{i_l } {G_{i_l ,s} } |^2  \
cdot g_l (\omega _{s\sigma } ) \cdot N_l (\omega _{s\sigma } ) + 
|\sum\limits_{i_r } {G_{i_r ,s} } |^2  \cdot g_r (\omega _{s\sigma } ) 
\cdot N_r (\omega _{s\sigma } )} \over {|\sum\limits_{i_l } 
{G_{i_l ,s} } |^2  \cdot g_l (\omega _{s\sigma } ) + 
|\sum\limits_{i_r } {G_{i_r ,s} } |^2  \cdot g_r (\omega _{s\sigma } )}}
\label{eq9.2}
\end{eqnarray}

for 3-polar contact 

\begin{eqnarray}
n_s=(|\sum\limits_{i_l }{G_{i_l ,s}}|^2\cdot g_l 
(\omega _{s\sigma})\cdot N_l(\omega_{s\sigma}) + 
|\sum\limits_{i_u}{G_{i_u ,s}}|^2\cdot g_u (\omega_{s\sigma}) 
\cdot N_u (\omega_{s\sigma }) + \nonumber \\
+|\sum\limits_{i_d }{G_{i_d ,s}}|^2\cdot g_d 
(\omega_{s\sigma})\cdot N_d(\omega_{s\sigma} ))/ \nonumber \\
/(|\sum\limits_{i_l}{G_{i_l,s}}|^2\cdot g_l(\omega_{s\sigma})+ 
|\sum\limits_{i_u}{G_{iu,s}}|^2\cdot g_u(\omega_{s\sigma}) + 
|\sum\limits_{i_d}{G_{i_d ,s}}|^2  \cdot g_d (\omega _{s\sigma} ))
\label{eq9.3}
\end{eqnarray}

And for 4-polar contact

\begin{eqnarray}
n_s=(|\sum\limits_{i_l }{G_{i_l ,s}}|^2\cdot g_l(\omega_{s\sigma}) 
\cdot N_l (\omega _{s\sigma } ) + |\sum\limits_{i_r } {G_{i_r ,s}} 
|^2 \cdot g_r(\omega_{s\sigma})\cdot N_r(\omega_{s\sigma})+ \nonumber \\ 
+|\sum\limits_{i_u }{G_{i_u ,s}}|^2\cdot g_u 
(\omega_{s\sigma})\cdot N_u(\omega_{s\sigma})
+|\sum\limits_{i_d }{G_{i_d ,s}}|^2  
\cdot g_d(\omega_{s\sigma}) 
\cdot N_d(\omega_{s\sigma}))/ \nonumber \\
/(|\sum\limits_{i_l }{G_{i_l,s}}|^2  
\cdot g_l (\omega_{s\sigma})+|\sum\limits_{i_r}
{G_{i_r ,s}}|^2\cdot g_r(\omega_{s\sigma})+
|\sum\limits_{i_u}G_{iu,s}|^2  
\cdot g_u(\omega_{s\sigma})+ \nonumber \\
+|\sum\limits_{i_d }{G_{i_d,s}}|^2  
\cdot g_d(\omega_{s\sigma})) 
\label{eq9.4}
\end{eqnarray}

\begin{eqnarray}
\partial Q/\partial t=2\pi\sum\limits_{s,\sigma}\left|G_{ls}
\right|^2\left|G_{rs}\right|^2\omega_{s\sigma}
g_l(\omega_{s\sigma})g_r(\omega_{s\sigma})\times \nonumber \\
\times\frac{N_l(\omega_{s\sigma})-N_r(\omega_{s\sigma})}
{\left|G_{ls}\right|^2g_l(\omega_{s\sigma})+
\left|G_{rs}\right|^2g_r(\omega_{s\sigma})
} 
\label{eq9.5}
\end{eqnarray}

\begin{eqnarray}
\lambda^{'}= L\left| {\frac{{\partial Q/\partial t}}{{T_r  - T_l }}} \right|
\label{eq9.6}
\end{eqnarray}

Approximation

\begin{eqnarray}
N_l (\omega _{s\sigma } ) - N_r (\omega _{s\sigma } ) 
\approx {{\partial N(\omega _{s\sigma } )} \over {\partial T}}(T_r  - T_l )
\label{eq9.7}
\end{eqnarray}

\begin{eqnarray}
\lambda^{'}(L)=2\pi L\sum\limits_{s,\sigma}\left|G_{ls} 
\right|^2\left|G_{rs}\right|^2\omega_{s\sigma}g_l 
(\omega_{s\sigma})g_r(\omega_{s\sigma})\times \nonumber \\
\times\frac{\partial N(\omega _{s\sigma })
/\partial T}{\left|G_{ls}\right|^2g_l(\omega_{s\sigma})+ 
\left|G_{rs}\right|^2 g_r(\omega_{s\sigma})} 
\label{eq9.8}
\end{eqnarray}

\section{ Phonon mean free path and size dependences } 

\begin{eqnarray}
\lambda ^{'} (l_{ph} ) = {{2\pi l_{ph} } \over {T^2 }}\sum\limits_{s,\sigma } 
{\left| {G_{ls} } \right|^2 \left| {G_{rs} } \right|^2 
\omega _{s\sigma }^2 g(\omega _{s\sigma } ){{N(\omega _{s\sigma } )
(N(\omega _{s\sigma } ) + 1)} \over {\left| {G_{ls} } \right|^2  + 
\left| {G_{rs} } \right|^2 }}} 
\label{eq10.1}
\end{eqnarray}

\begin{eqnarray}
G_{rs}  = \sum\limits_{i_r } {G_{i_r } C_{i_r s} } 
\label{eq10.2}
\end{eqnarray}

 The mean free path along the tube lph.

\begin{eqnarray}
l_{ph}  \approx \sqrt {{S \over N}}  = \sqrt {{{l_d 
\cdot l_{ph} } \over N}},\hspace{1cm}l_{ph}  \approx {{l_d } \over { < N > }}
\label{eq10.3}
\end{eqnarray}

\begin{eqnarray}
<N> = {1 \over \Delta }\int {n(\omega )g(\omega )}d\omega ,
\hspace{1cm}l_{ph} \sim{1 \over T}
\label{eq10.4}
\end{eqnarray}
\newpage
.

\special{em:graph fig_10_1.gif}
\vspace{11.58cm}
{ Fig.10.1. One dimensional phonon dynamics. Calculated total thermal 
conductivity length dependence that includes all vibration branches of armchair 
NT. Lph is phonon mean free path, a is C- bond length.  }

\newpage
\section{ Thermal coefficient temperature dependence} 

\begin{eqnarray}
\lambda _0  = {{2\pi aG_0^2 } \over {\hbar \omega _0 }}k
\label{eq11.1}
\end{eqnarray}

\vspace{0.5cm}
\special{em:graph fig_11_1.gif}
\vspace{10.74cm}
{ Fig.11.1 . Thermal conductivity temperature dependence. Armchair NT. Curve 1 
corresponds to z-branch contribution, curve 2 to j-branch and r-branch 
contribution is presented by curve 3.  }

\newpage
.

\special{em:graph fig_11_2.gif}
\vspace{10.26cm}
{ Fig.11.2 Thermal conductivity temperature dependence. Zig-zag NT. Curve 1 
corresponds to z-branch contribution, curve 2 to j-branch and r-branch 
contribution is presented by curve 3.  }

\section{XII.           Generalized phonon transport equation} 
 
\begin{eqnarray}
\dot W_{l \to s\sigma }  = {{2\pi } \over \hbar }\left| {G_{ls} } 
\right|^2 g_l (\omega _{s\sigma } )N_l (\omega _{s\sigma } )
(1 + n(\omega _{s\sigma } ))
\label{eq9a.1}
\end{eqnarray}

\begin{eqnarray}
{{\partial n_{s\sigma } } \over {\partial t}} = D_z^{s\sigma } 
{{\partial ^2 n_{s\sigma } } \over {\partial z^2 }}
\label{eq12.1}
\end{eqnarray}

\begin{eqnarray}
D_z^{s\sigma }  = {{2\pi l_{ph} } \over \hbar }\left| 
{G_{ls\sigma } } \right|^2 g(\omega _{s\sigma } )
\label{eq12.2}
\end{eqnarray}

\begin{eqnarray}
G_{ls\sigma }  = G_{rs\sigma }  = \sum\limits_{i_l } 
{G_{i_l } C_{i_l s\sigma } } 
\label{eq12.3}
\end{eqnarray}

 Temperature distribution along the tube

\begin{eqnarray}
T(z) = \omega _{s\sigma } /Ln\left( {1 + {L \over {n_l (\omega _{s\sigma } ) + 
[n_r (\omega _{s\sigma } ) - n_l (\omega _{s\sigma } )]z}}} \right)
\label{eq12.4}
\end{eqnarray}

\newpage
.

\special{em:graph fig_12_1.gif}
\vspace{2.93cm}
{ Fig.12.1 Phonon jumps over a carbon armchair structure (arrows). Nanotubes are 
ideal one-dimensional systems as to the phonon dynamics. }

\vspace{0.5cm}
\special{em:graph fig_12_2.gif}
\vspace{11.32cm}
{ Fig.12.2. One dimensional phonon dynamics. Exact temperature distribution 
along the tube given by (4.4). Tl  ,Tr.are temperatures of baths connecting with 
tube ends, L is the tube length. End points come together all modal (partial) 
temperatures for each ss. }

\newpage
\section{Summary and discussion} 

\vspace{0.5cm}
\special{em:graph fig_13_1.gif}
\vspace{4.12cm}
{ Fig.13.1. Averaging procedures and speed of propagation. a is interatomic 
distance, <x> is mean atomic shift , x is extended coordinate of the freedom's 
degree, V is excitation velocity, m is atom mass, k is elasticity coefficient. }

\vspace{0.5cm}
\special{em:graph fig_13_2.gif}
\vspace{7.49cm}
{ Fig.13.2. Temperature dependence of thermal conductivity coefficient. Two 
competitive tendencies. }

The boundary conditions for microsound spreading through periodic 
NT complexes
\begin{eqnarray}
\left\{ \matrix{
0.75ka{{\partial  < z > } \over {\partial z}} + ( - 1)^{l,r} m\omega ^2  
< z >  = const \hfill \cr 
{{\partial  < z > } \over {\partial z}} = const \hfill \cr}  \right.
\label{eq13.1}
\end{eqnarray}

\begin{eqnarray}
< z >  = Ae^{ifz}  + Be^{ - ifz} 
\label{eq13.2}
\end{eqnarray}

 A complex approach PQDM was applied to describe phonon dynamics and heat 
spreading processes in carbon NT-solid medium composite material and in 
nanotubes embraced by a polymer molecule. 

 Atomic vibration dynamics was considered for carbon zero-hirality nanotubes 
connected with solid matrix. Vibrational eigenmodes, density of states and 
amplitude distribution for tubes inserted into a cylindrical well inside the 
medium and bounded with its hard inner wall were calculated. Phonon band 
parameters were analyzed in linear approximation for three types of vibration: 
athimuthal or tangential j-branch, radial breathing mode (r-branch) and 
longitudinal z-branch.

 It's shown that phonon propagation in actual nanotubes is characterized by a 
kind of "compactification" of circular freedom's degree due to the big phonon 
mean free path. Nanotubes of actual diameters are ideal one-dimensional phonon 
qnd heat conductors.

 Phonon band structure was investigated for armchair nanotubes on the base of 
hierarchical law and system symmetry.

 Thermal fluxes and thermal conductivity were considered in PQDM. Temperature 
dependences were obtained. The mechanism of heat conductivity high temperature 
damping is cleared. Two competitive tendencies produce thermal conductivity 
maximum at intermediate temperatures (100-300)K.

 The exact solution of generalized thermal conductivity equation was obtained 
for nanotubes. Temperature distribution along the tube was derived analytically. 

 Size dependences were considered for thermal conductivity. It was shown the 
linear increasing of heat conductivity with the growth of the phonon mean free 
path.

 The explanation of the nature of good thermal conductivity in carbon and 
carbon-like materials by existing of the soft vibration branch (low frequency 
RBM phonons with high DOS at thermal energies) accompanied by structure hardness 
(high frequency j- and z-branches) providing big mean free path for phonons 
(Fig.5.2). 

 Adding of new layers or new walls to single-walled NT makes breathing r-branch 
of vibrations harder and causes the sharp decreasing of phonon density of states 
at the same phonon mean free path

 Pressure decreases thermal conductivity. The effect is connected with total 
hardening of all bonds and phonon modes going away from active thermal zone.

 Melting decreases thermal conductivity by another reason: the phonon's mean 
free path becomes small. 

 Modified Fourier equation of heat propagation in microscopic molecular channel 
was obtained.

 Boundary problem for temperature distribution in channel-medium system was 
considered in cylindrical symmetry. The method of partial solution building for 
cylindrical boundary problem was proposed. Temperature distribution isotherms 
were calculated.

 Concentration dependences were considered for thermal conductivity of NT-solid 
matrix composite. It was shown the strongly non-linear increasing of effective 
heat conductivity with the growth nanotubes concentration.

 Phonon dynamics was considered for carbon zero-chirality nanotubes embraced by 
polymer molecule. Vibrational eigenmodes, density of states and amplitude 
distribution for a bounded tube were calculated. Phonon band parameters were 
obtained in linear approximation for three types of vibration.

 It was shown that insertion of armchair nanotube inside a medium or its coating 
by polyacetilene molecule considerably changes the structure of radial breathing 
phonons

\end{document}